\documentclass[prl,aps,reprint,superscriptaddress,showpacs,longbibliography]{revtex4-1}

\usepackage{amsmath} \usepackage{amssymb} \usepackage{graphicx}
\usepackage{xcolor} \usepackage{subfigure} \usepackage{color}

\begin{document}

\title{ Topological quantum control: Edge currents via Floquet
  depinning of skyrmions in the $\nu = 0$ graphene quantum Hall
  antiferromagnet}

\author{Deepak Iyer} \affiliation{Department of Physics \& Astronomy,
  Bucknell University, Lewisburg, Pennsylvania 17837, USA}
\author{Matthew S. Foster} \affiliation{Department of Physics and
  Astronomy, Rice University, Houston, Texas 77005, USA}
\affiliation{Rice Center for Quantum Materials, Rice University,
  Houston, Texas 77005, USA}

\date{\today}

\begin{abstract}
  We propose a defect-to-edge topological quantum quench protocol that
  can efficiently inject electric charge from defect-core states into
  a chiral edge current of an induced Chern insulator.  The initial
  state of the system is assumed to be a Mott insulator, with
  electrons bound to topological defects that are pinned by disorder.
  We show that a ``critical quench'' to a Chern insulator mass of
  order the Mott gap shunts charge from defects to the edge, while a
  second stronger quench can trap it there and boost the edge
  velocity, creating a controllable current.  We apply this idea to a
  skyrmion charge in the $\nu = 0$ quantum Hall antiferromagnet in
  graphene, where the quench into the Chern insulator could be
  accomplished via Floquet driving with circularly polarized light.
\end{abstract}

\maketitle


The pervasive emphasis on topology in modern condensed matter physics
is due to two seemingly disparate pursuits.  On one hand, topology can
``protect'' interesting quantum phenomena against imperfections or
environmental decoherence, enabling ballistic propagation through edge
states \cite{haldane88,hasan10,bernevig13}, and quantum information
storage and manipulation \cite{nayak08}.  On the other hand,
topological defects (solitons and instantons \cite{rajaraman82}) in
effective quantum field theories play a central role in
non-perturbative frameworks for strongly correlated electron systems
\cite{fradkin13,altland10}.  In particular, topological defects in
bosonic order parameters can bind fermion quasiparticles in their
cores; the topological charge of the defect then determines the
ground-state electrical charge or current induced by it
\cite{goldstone81,jaroszewicz84,carena90}.  Applications of this
principle in condensed matter physics range from producing
fractionally charged zero modes in graphene \cite{hou07}, to novel
mechanisms for superconductivity \cite{grover08} and quantum
criticality \cite{tanaka05,fu11,goswami14,liu17}.

\begin{figure}[!b]
  \centering
  \includegraphics[width=0.4\textwidth]{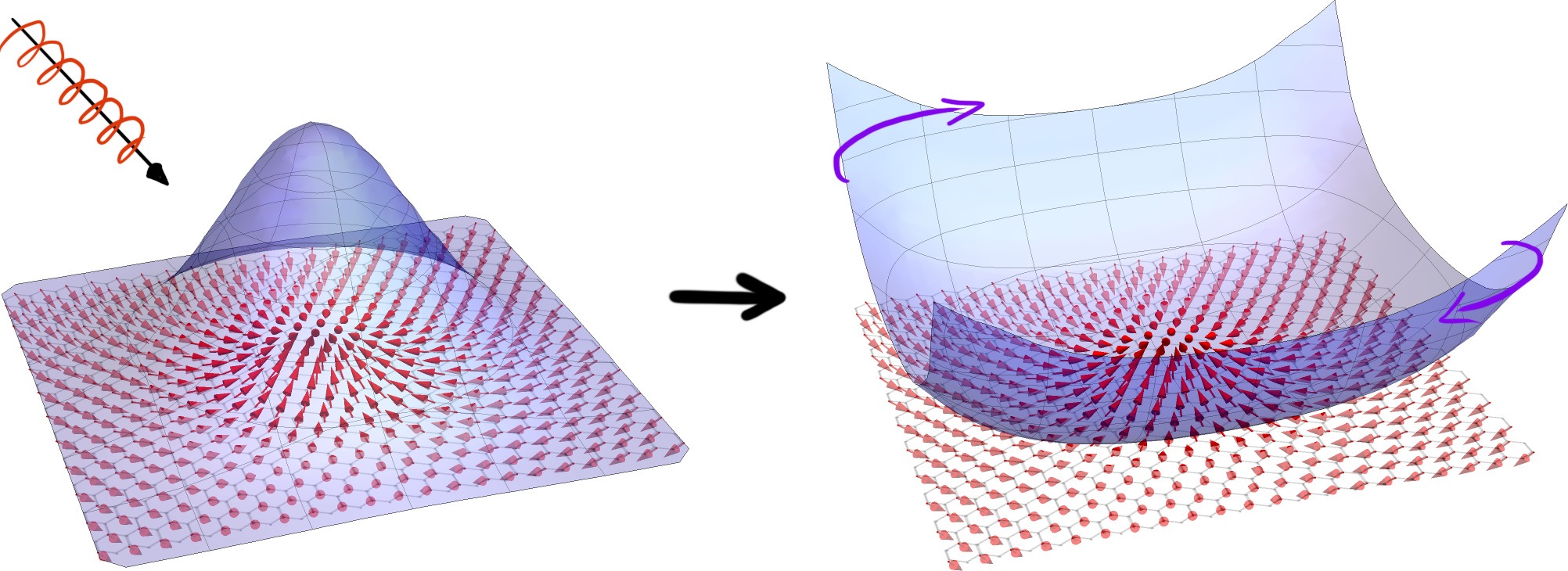}
  \caption{A topological quantum control protocol for creating chiral
    edge currents.  A Chern insulating (Haldane) mass is induced by
    the application of circularly polarized light
    \cite{oka08,kitagawa11,gu11,kundu14,torres14,dehghani15,cavalleri20},
    incident on Mott-insulating graphene with a single skyrmion
    embedded in a N\'eel antiferromagnetic texture. A single hole is
    doped into the negative-energy core state bound to the skyrmion
    \cite{jaroszewicz84,carena90,goswami14,liu17}.  Such a defect can
    arise as a pinned charge carrier in the $\nu = 0$ quantum Hall
    state of graphene \cite{jung09,kharitonov12,young14,jolicoeur19}.
    When the strength of the incident light is at criticality (with
    respect to the topological transition induced by a Haldane mass),
    the doped hole ballistically migrates to the edge of the sample. A
    second quench (not shown) deep into the topological phase sets up
    a circulating current of the charge, as indicated by the purple
    arrows. The spins show the skyrmion texture on the underlying
    honeycomb lattice and the blue surface illustrates the charge
    density. For weak and moderate quenches, the spin texture in our
    mean-field calculation is not significantly scrambled.}
  \label{fig:illustration}
\end{figure}

\begin{figure*}[th!]
  \centering
  \includegraphics[width=7cm]{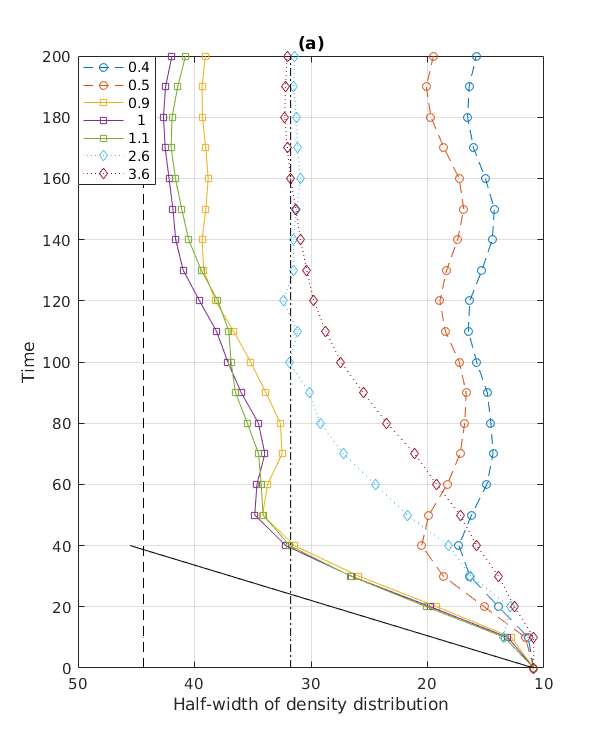}
  \includegraphics[width=7cm]{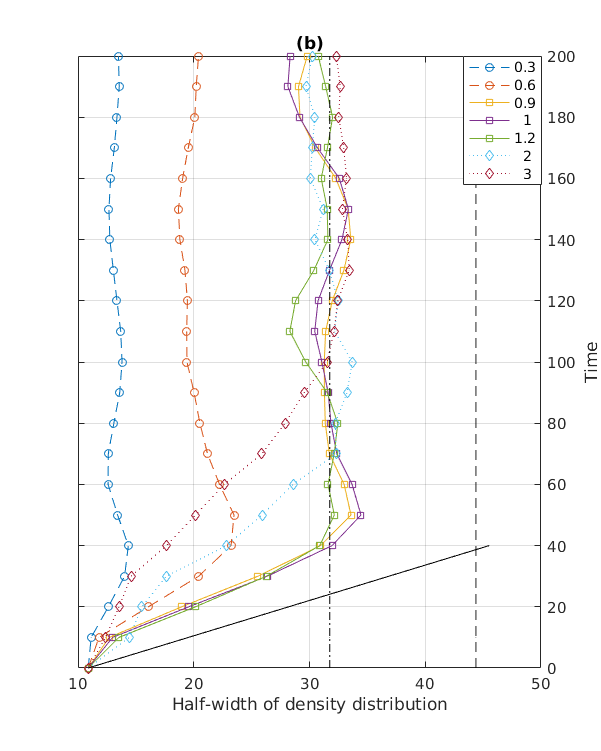}
  \caption{Half-width (standard deviation) of the electric charge
    distribution around the center of the skyrmion texture as a
    function of time.  A single hole is doped into the negative-energy
    core state of the skyrmion defect in a N\'eel background before
    the quench. This hole is ``depinned'' by the quenching on of a
    competing mass (order parameter).  (a) corresponds to a quench
    into the Chern insulating phase of the Haldane model, and (b)
    corresponds to a quench into a model with a sublattice-staggered
    mass [charge density wave (CDW) order] that displays no
    topologically nontrivial phase.  The legend indicates values of
    the quench coupling strength normalized so that the equilibrium
    gap (with coexisting N\'eel and Chern insulator or CDW orders)
    closes at unity: for (a) it is $3\sqrt{3}t_{2}/m_{N}$ and for (b)
    $m_{C}/m_{N}$.  Here $t_2$ ($m_C$) is the Haldane hopping (CDW)
    strength, and $m_N$ is the energy gap in the homogeneous N\'eel
    background.  In both figures, the solid black line represents the
    light-cone given by the Fermi velocity, $v_{F} = \sqrt{3}t/2$, the
    black dash-dot line represents the value of the half-width for a
    uniform charge distribution
    across the graphene flake, and the black dashed line represents
    the value for all charge concentrated on the edge.  As the
    strength of the Haldane coupling $t_2$ increases, the initial
    charge density profile spreads faster and faster until the gap
    closes, after which the spreading slows down. Note that curves
    with values of the half-width greater than the uniform value
    (black dot-dash line) have greater charge density on the
    edge/boundary of the sample than the bulk.  Near criticality in
    (a), edge states are strongly populated, and expansion is nearly
    ballistic (modulo boundary effects and a slow start) at the Fermi
    velocity.  However, we do not see any corresponding shift of
    charge to the edge in the CDW case due to the absence of
    topologically nontrivial edge states in (b).  Thus, there is a
    nontrivial effect from pumping the sample with circularly
    polarized light to create the Haldane mass (a).}
  \label{fig:stdevcharge}
\end{figure*}

In this Letter, we combine these two notions of topology in a
dynamical setting, using a quantum quench.  We consider a quench into
a 2D Chern insulator (Haldane) phase \cite{haldane88}, with
dynamically generated edge states in a honeycomb model with Dirac
fermions. Previous work considered quenches from spatially
homogeneous, trivial or topological initial states
\cite{foster13,foster14,dalessio15,liao15,caio15,wang16,caio16,liou18}.
Here, we instead envision an antiferromagnetic (N\'eel-ordered) Mott
insulating initial state, wherein topological skyrmion defects trap
Dirac electrons or holes in their cores
\cite{jaroszewicz84,carena90,grover08,herbut12,goswami14,liu17}.  The
skyrmion defects are initially pinned by disorder, providing a robust
reservoir of localized charge for a weakly-doped initial Mott
insulating state. We subject the Mott state to a quenching on of the
Chern insulator mass, which competes with the N\'eel order. We find
that a ``critical'' quench can efficiently transfer the charge bound
to a skyrmion defect to the quench-induced chiral edge states formed
at the boundary of the sample.  Here, a critical quench means that the
induced Haldane mass is tuned to match the ground state Mott gap.
After transferring the charge to the edge, a second quench deeper into
the Haldane phase traps the charge at the sample boundary, inducing a
tunable circulating edge current. Thus our double-quench protocol
provides a way to efficiently depin electric charge from topological
defects of the Mott reservoir and transfer it to the boundary of a
dynamically-induced topological insulator.

Our protocol could be implemented using Floquet driving with
circularly polarized light in graphene \cite{oka08,kitagawa11}, as
explicated in Fig.~\ref{fig:illustration}.  Many theoretical works
have examined Floquet-induced Chern insulator states that could be
realized in semimetallic graphene, with proposals to measure induced
topological edge states through conducting leads
\cite{oka08,kitagawa11,gu11,kundu14,torres14}.  However, even in the
ideal case of ballistic Landauer transmission through such a driven
device, the complicated matching conditions between time-dependent
device and equilibrium contact states typically leads to the
prediction of a non-quantized response in transport
\cite{gu11,kundu14,torres14}.  Since the Dirac point conductivity of
graphene itself is always of order $e^2/h$ at low temperature
\cite{graphenerev}, the transition to almost-quantized edge transport
is not easy to verify in a two-terminal Landauer geometry.  Transport
measurements typically require long averaging times, while
steady-state illumination on ungapped graphene can excite many hot
electron-hole carriers. Then, it becomes essential to model the
distribution function of the electrons induced by the drive
\cite{dehghani15}, which goes beyond the Floquet-Landauer theory.
Ultrafast pumping and detection can mitigate these limitations, see
Ref.~\cite{cavalleri20} for a recent experiment.

In this Letter, we instead propose to apply a circularly-polarized
Floquet drive to the \emph{strongly insulating} $\nu = 0$ quantum Hall
state of graphene \cite{checkelsky08,young12}.  The insulating gap in
the $\nu = 0$ state arises due to quantum Hall magnetism
\cite{girvin99} in the nearly SU(4)-symmetric zeroth lowest Landau
level (LLL) \cite{abanin06,nomura06,yang06}.  Due to the LLL
projection, the charge carriers in quantum Hall magnets typically
exhibit a topological skyrmion texture in the magnetic order
\cite{girvin99,sondhi93,arovas99}.  This $\nu=0$ state in graphene has
only recently been understood as a quantum Hall \emph{anti}ferromagnet
(AFM) in physical spin \cite{jung09,kharitonov12,young14}, which
should support SU(2) skyrmion charge excitations
\cite{yang06,jolicoeur19}.  A similar setup could start with Mott
states in twisted bilayer graphene systems \cite{cao18,lu19,sharpe19}.

When a quantum Hall magnet is doped very slightly away from integer
filling, skyrmionic charge carriers are expected to be pinned randomly
throughout the sample by weak quenched disorder \cite{girvin99}. We
model the $\nu = 0$ state of graphene using a honeycomb lattice model
subject to mean-field N\'eel antiferromagnetism, with a single
skyrmion defect localized in the center of a finite square sample
(with open boundary conditions).  The defect traps a single pair of
subgap, positive and negative-energy core states.  The core states
form a particle-hole-symmetric pair in the $K$ and $K'$ valleys
\cite{goswami14}.  We consider a system tuned just below half-filling,
with the negative-energy core occupied by a hole. The Floquet drive is
treated as an instantaneous quantum quench that turns on the Haldane
Chern insulator mass \cite{dalessio15,caio15,wang16,caio16,liou18}, an
approximation \cite{kitagawa11} that becomes exact in the
high-frequency limit \cite{magnus54,casas01}.

\begin{figure*}[htb]
  \centering
  \includegraphics[width=0.49\textwidth]{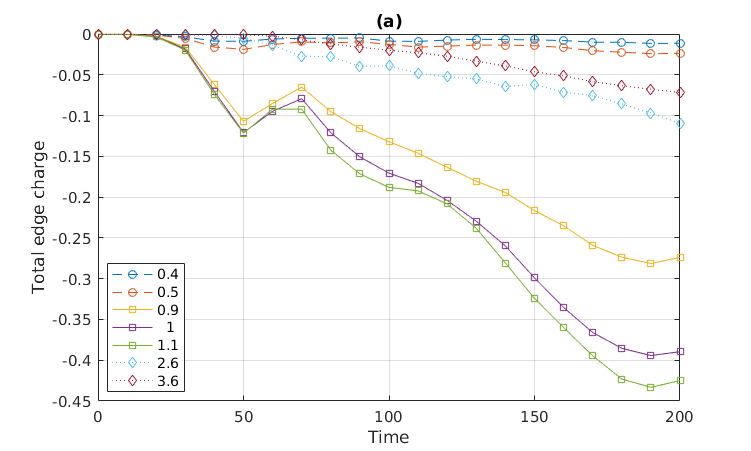}
  \begin{picture}(0,0)
    \put(-189,18){\includegraphics[width=3.5cm]{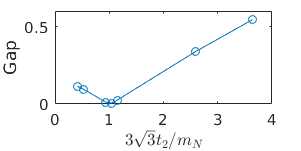}}
  \end{picture}
  \includegraphics[width=0.49\textwidth]{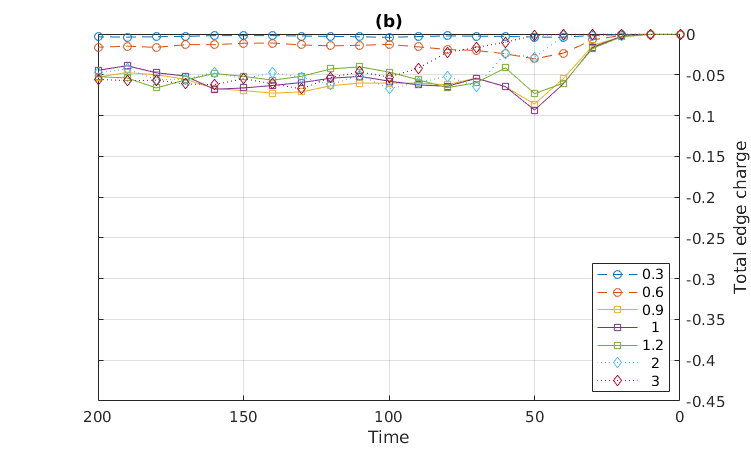}
  \begin{picture}(0,0)
    \put(-189,18){\includegraphics[width=3.5cm]{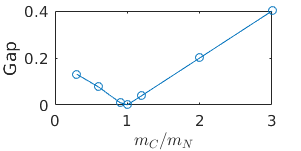}}
  \end{picture}
  \caption{Total charge on the boundary (measured relative to the
    half-filled background) following the same quench pictured in
    Fig.~\ref{fig:stdevcharge}.  (a) corresponds to the case when we
    quench into the Haldane model, (b) the quench into a topologically
    trivial CDW model.  Legends indicate the relevant quench parameter
    (see Fig.~\ref{fig:stdevcharge}). In (b), we see no
    significant motion of charge to the boundary, whereas in the
    Haldane model (a), almost half of the total charge has moved to the
    boundary. Note that from Fig.~\ref{fig:stdevcharge} it appears
    that almost all the charge has moved to the boundary (see black
    dashed line). Here, we compute the charge at the very edge of the
    sample. If we include a thicker boundary layer, then we capture
    most of the electric charge initially bound to the skyrmion-core.
    In both cases, far from the critical point, we don't see
    significant motion of charge to the boundary. Insets in each
    figure show the equilibrium gaps as a function of the relevant
    tuning parameter.}
  \label{fig:edge_charge}
\end{figure*}

\begin{figure}[htb]
  \includegraphics[width=0.48\textwidth]{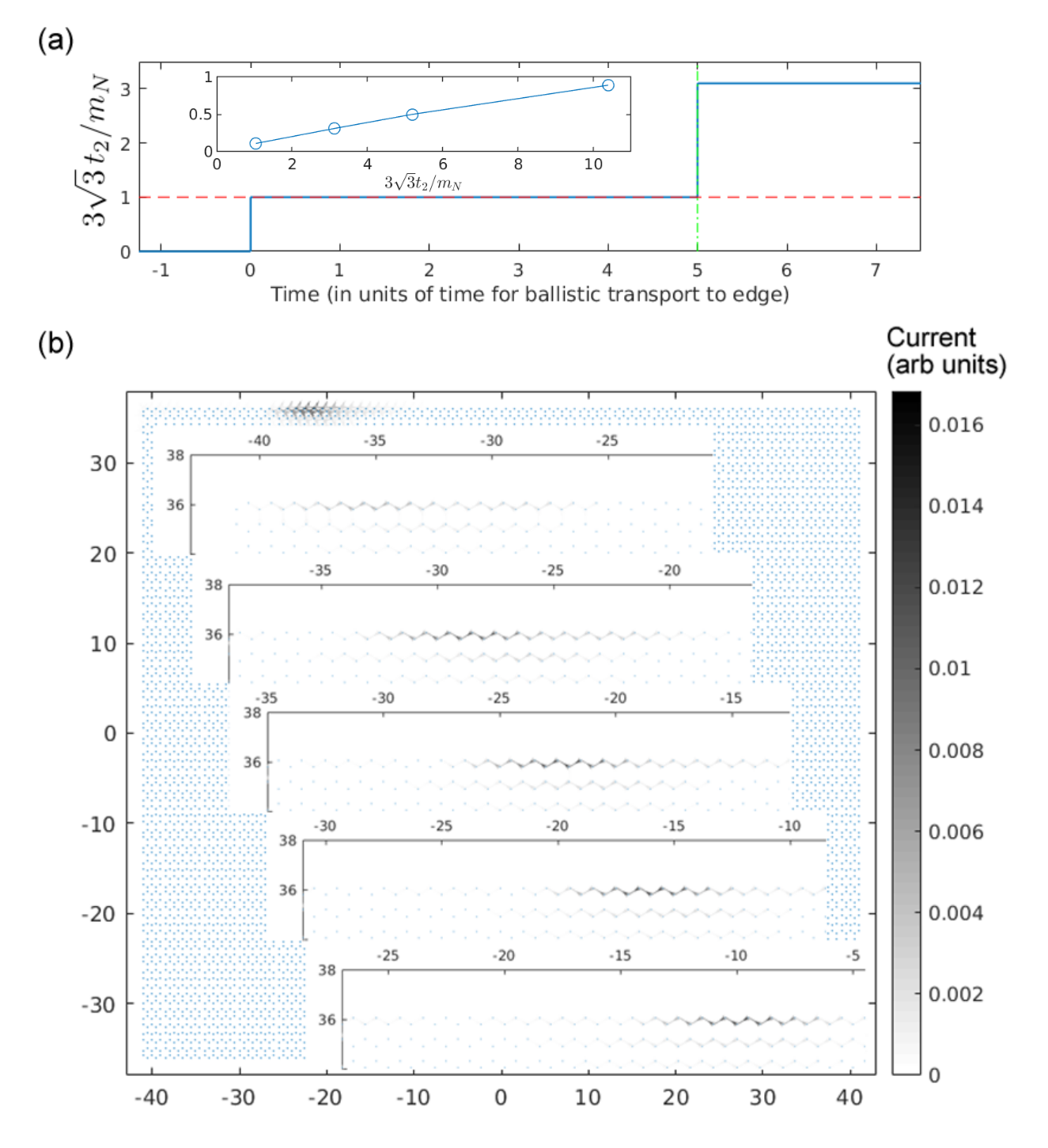}
  \caption{Edge currents induced by a double quench.  The
    double-quench protocol is shown in (a), the currents in (b).  The
    inset in (a) shows the average edge velocity as a function of the
    Haldane mass after the second quench, in units of the lattice
    parameter $\times$ hopping energy; the blue line is a guide to the
    eye.  The edge velocity increases with increasing strength of the
    second quench, allowing control of the induced edge current. The
    enlarged snapshots in (b) indicate time evolution going down along
    the images.  The current shifts towards the right as time evolves
    (note horizontal axis). In each snapshot, the arrows point in the
    direction of the charge current; charge and particle currents are
    opposite for the edge-migrated hole.  The background image shows
    the full sample at a particular time snapshot.  Note that there is
    almost no current anywhere else in the sample. The scale on the
    right indicates the value of the current.}
  \label{fig:edge_currents}
\end{figure}

We show that a quench of critical strength (matching the dynamically
generated Haldane and ground-state AFM mass gaps) depins the doped
hole, which ballistically propagates to edge. To demonstrate that this
effect is a nontrivial consequence of the Chern mass quench, we show
that a similar ``sticking'' of the depinned charge to the sample
boundary does not occur for a quench employing the topologically
trivial charge density wave mass; see Figs.~\ref{fig:stdevcharge} and
\ref{fig:edge_charge}.  Finally, we show that a second quench that
intensifies the Haldane mass can trap the migrated charge at the
boundary of the sample for a long time, inducing a circulating
electrical current with an edge velocity determined by the drive
strength (Fig.~\ref{fig:edge_currents}).  The key new ingredient in
our nonequilibrium scheme is that we controllably induce and populate
the edge states, depinning charges from topological defects in a
strongly insulating prequench state.


\textit{Model.}---We consider spin-1/2 electrons hopping on a
honeycomb lattice, subject to an inhomogeneous, sublattice-staggered
Zeeman field that mimics a skyrmion defect texture in the N\'eel
background.  The prequench Hamiltonian is given by
\begin{multline}
  \label{eq:hamskyr}
  H_{\text{sk}} = - t \sum_{\alpha}\sum_{r,s} \left(
    c^{\dag}_{r,\alpha} c_{r+s,\alpha} + \text{H.c.}  \right) +
  \\
  + m_{N} \sum_{r}(-1)^{\tau} \vec{n}(r)\cdot c^{\dag}_{r,\alpha}
  \vec{\sigma}^{\alpha\beta} c_{r,\beta}.
\end{multline}
The first term encodes nearest-neighbor hopping; H.c.\ denotes the
Hermitian conjugate.  The second term is the Zeeman coupling, where
$\tau$ is 0 for $r\in A$ and 1 for $r\in B$ ($A,B$ denote the
sublattices).  Here $\alpha$ is the spin index, and $\vec{\sigma}$ is
a vector formed from the Pauli matrices acting on physical spin.

In Eq.~(\ref{eq:hamskyr}), we choose $\vec{n}(r)$ to be a skyrmion
texture in the continuum \cite{rajaraman82}. On the lattice, we
superimpose this onto the vertices with the skyrmion centered on a
plaquette at the center of the lattice.  The texture is parameterized
by $\omega(z) \equiv z/\lambda$, where $\lambda$ is a scale that
determines the core size, and $z = x + i y$.  The spin texture is
given by
$ n_{z} = \left(|\omega|^2 - 1\right)/\left(|\omega|^2 + 1\right) $,
$ n_{x} + i n_{y} = \left(2\omega\right)/\left(|\omega|^2 + 1\right).
$ For $|z| \gg \lambda$ and $|z|\ll\lambda$, the texture becomes
uniform and points in the $\pm z$ direction (i.e., out of the graphene
plane) and is essentially a homogeneous N\'eel mass, $m_{N}$. This
Hamiltonian has a nonzero gap, and a particle-hole symmetric pair of
positive- and negative-energy bound states, localized to the skyrmion
core \cite{goswami14,liu17}.  The spin textures of the core states
wind like the original skyrmion texture shown in
Fig.~\ref{fig:illustration}, since the Zeeman term pins the spin
orientation.  Each core state can accommodate a single electron
charge; the skyrmion carries unit topological or Pontryagin charge
\cite{rajaraman82,kharitonov12}.  We fill all the states at negative
energy, \emph{except} for the negative-energy core state (which is
doped with a hole).  Equivalently, we could fill all negative energy
states and the positive-energy core state.  The ground-state density
profile reflects the missing charge bound to the core state. By
contrast, at half filling, the charge density would be uniform.

The quantum quench evolves this initial state with an additional
Haldane term, $H = H_{\text{sk}} + H_{H}$, where
\begin{equation}
  \label{eq:haldane}
  H_{H} 
  = 
  i t_{2}
  \sum_{\alpha}\sum_{r,s\in\{\text{nnn}\}} 
  c^{\dag}_{r,\alpha} c_{r+s,\alpha} 
  + 
  \text{H.c.},
\end{equation}
such that the positive signs form counterclockwise triangles in each
sublattice \cite{haldane88}.  In order to correlate any unique
behavior that emerges from quenching into a topologically nontrivial
phase, we also carry out a quench where we turn on a topologically
trivial charge density wave (CDW) potential (equivalently, a
sublattice-staggered mass), $H = H_{\text{sk}} + H_{C}$, where
\begin{equation}
  \label{eq:cdw}
  H_{C} 
  = 
  m_{C}\sum_{r,\alpha} 
  (-1)^{\tau} 
  c^{\dag}_{r,\alpha} c_{r,\alpha}.
\end{equation}
Here, as before, $\tau=\{0,1\}$ for $r\in \{A,B\}$ respectively. All
states of this model are topologically trivial.  In particular, there
are no current-carrying edge states.  Both the Chern-insulator and CDW
terms compete with the N\'eel order; sufficiently large $t_2$ or $m_C$
relative to $m_N$ can close the bulk gap in equilibrium. We note that
in a finite sample as studied numerically here, both models
$H = H_{\text{sk}} + H_{H,C}$ show edge states \emph{near gap closure}
coming from the open boundary conditions.


\textit{Numerics.}---In equilibrium, the model with a homogeneous
N\'eel mass undergoes a quantum phase transition at
$m_{N} = 3\sqrt{3}t_{2}$, at which point the gap closes, and edge
states appear \cite{haldane88}.  As $t_{2}$ increases beyond this
value, a gap reappears. However, edge states that connect across the
gap persist.  For the model with the skyrmion texture, we observe that
the gap closes at approximately the same point [see inset,
Fig.~\ref{fig:edge_charge}(a)], with the subtlety that the two
skyrmion-core bound states are present almost until the gap closes. We
tune the quench through this quantum phase transition and study the
resulting evolution of the charge density and charge current. Since
the Haldane term is diagonal in spin, we do not expect any nontrivial
behavior in the spin sector. We perform calculations on a
square-shaped section of the honeycomb lattice, with two zigzag edges
and two armchair edges using numerical exact diagonalization on a
lattice 82 plaquettes wide. The specific shape of the flake
is irrelevant since our initial spatially inhomogeneous Hamiltonian
breaks lattice symmetries.


\textit{Charge dynamics.}---The first observation and primary result
is the rapid population of the edge of the sample near
criticality. The skyrmion-core (hole) charge density starts out near the
center of the sample, but rapidly moves outwards, as seen in
Figs.~\ref{fig:stdevcharge} and \ref{fig:edge_charge}.  In
Fig.~\ref{fig:stdevcharge}, the radial spread of the charge density
(standard deviation) is plotted against time for different values of
the Haldane mass $t_{2}$.  Away from the critical point,
$3\sqrt{3}t_{2}/m_{N} \approx 1$, the spreading is slow. In the
weak-quench regime, this is because of the absence of edge states and a nonzero gap. In
the strong-quench regime, edge states are present but have weak
overlap with the initial state. It is only near criticality, i.e.\ where the
gap closes and edge states appear, that we get rapid charge accumulation at the edge.  The
small gap between the bulk-core and edge states promotes efficient
hybridization between these. It is nevertheless interesting to note
that the charge density preferentially occupies the edge as opposed to
a more uniform distribution.
As the system continues to evolve in time, we observe a slow
relaxation.

In comparison, as seen in Figs.~\ref{fig:stdevcharge}(b) and
\ref{fig:edge_charge}(b), the topologically trivial CDW quench does
not produce the saturation of charge at the sample boundary that we see in
the Haldane quench.  Although there are no topologically protected
chiral edge states in this case, almost degenerate edge states
nevertheless exist on account of the finite sample and appear at gap
closure at $m_{N} = m_{C}$ [in equilibrium, see inset,
Fig~\ref{fig:edge_charge}(b)].  In spite of this, we do not see
significant population of these edge states, indicating that the
topologically nontrivial nature of edge states that appear when we shine
circularly polarized light play a key role in the effect we observe.


\textit{Edge currents.}---The edge states of the Haldane model show
nontrivial currents circulating around the sample in the topological
phase, their sense depending on the sign of $t_{2}$ \cite{haldane88}.  
We therefore expect to see a circulating edge
charge when we quench into the topological phase, since the charge
initially bound to the skyrmion is pushed out to the edge.

In the equilibrium model, strong edge currents appear deep in the
topological phase.  When we quench on the Haldane coupling,
charge rushes to the boundary only near criticality. However, near
criticality, the edge currents are not strong (owing in part to the
shallow edge velocity).  Deeper in the topological phase, the edge
velocity is enhanced and edge currents are strong; however, a direct
deep quench does not shift significant charge to the boundary.

In order to generate a strong edge current,
we therefore adopt a \emph{double quench} protocol, see
Fig.~\ref{fig:edge_currents}(a).  The first quench shifts the charge
to the edge, and the second quench strongly confines this charge and
induces circulation with finite velocity [see inset,
Fig.~\ref{fig:edge_currents}(a)].  We illustrate this effect in
Fig.~\ref{fig:edge_currents}(b).  The large figure shows a single
temporal snapshot as an illustration.  The plot itself represents the
current flowing in each bond between the lattice sites.
The blown-up images show snapshots at increasing time after the second
quench (the first quench is allowed to evolve until we maximize the
edge charge). Note how the current shifts to the right i.e., a
clockwise flow. The arrows in each snapshot are opposite to the
overall flow
since this is a hole current.  Applying the double-quench protocol to
a sample with a finite density of hole-doped skyrmions would trap a
finite proportion of the core charge at the boundary, with the second
quench producing a tunable, steady-state edge current.


\textit{Conclusion.}---Charges pinned to topological defects in a Mott
insulator can serve as a charge reservoir that can be utilized to
efficiently load chiral edge states in a quantum-quench induced Chern
insulator, producing a tunable circulating edge current.  Such a
quench could be simulated in the $\nu = 0$ quantum Hall
antiferromagnetic state in graphene
\cite{jung09,kharitonov12,young14,yang06,jolicoeur19}, using a Floquet
drive \cite{oka08,kitagawa11}.  Future work could apply the same
strategy to twisted bilayer Mott states \cite{cao18,lu19,sharpe19}.

The induced edge current predicted here could be measured via the
laser-triggered photoconductive switch technique employed in the
experiment \cite{cavalleri20}. Alternatively, it might be possible to perform an all-optical measurement 
of the THz light re-emitted by a sufficiently large edge current induced 
by an intense THz pulse.


We thank Junichiro Kono and Chia-Chuan Liu for useful discussions.
We thank Yunxiang Liao for a collaboration on a precursor to 
this work.
This work was supported by U.S.\ Army Research Office Grant
No.~W911NF-17-1-0259, and by NSF CAREER Grant No.~DMR-1552327.  D.I.\
is grateful to the Rice Center of Quantum Materials for its
hospitality during various states of this work.

\end{document}